\documentstyle[aps]{revtex}

\begin{document}
\tightenlines
\title{Current density inhomogeneity throughout the thickness of superconducting
films and its effect on their irreversible magnetic properties}
\author{R. Prozorov$^1$, E. B. Sonin$^{2,3}$, E. Sheriff$^1$, A. Shaulov$^1$and Y.
Yeshurun$^1$}
\address{$^{1}$Institute of Superconductivity, Department of Physics, Bar-Ilan\\
University, 52900 Ramat-Gan, Israel}
\address{$^2$The Racah Institute of Physics, The Hebrew University of\\
Jerusalem, Jerusalem 91904, Israel \\
$^3$Ioffe Physical Technical Institute, St.~Petersburg 194021, Russia}
\date{February 15, 1998}
\maketitle

\begin{abstract}
We calculate the distribution of the current density $j$ in superconducting
films along the direction of an external field applied perpendicular to the
film plane. Our analysis reveals that in the presence of bulk pinning $j$ is
inhomogeneous on a length scale of order the inter vortex distance. This
inhomogeneity is significantly enhanced in the presence of surface pinning.
We introduce new critical state model, which takes into account the current
density variations throughout the film thickness, and show how these
variations give rise to the experimentally observed thickness dependence of $%
j$ and magnetic relaxation rate.
\end{abstract}

\draft

\section{Introduction}

The magnetic behavior of type-II superconductors depends strongly on the
sample shape \cite
{brandt96,gurevich94,brandt93,jooss96a,vlasko,jooss96,zeldov,norris}.
Significant progress has recently been made in understanding the effects of
a sample aspect ratio on its magnetic behavior\cite{brandt96}, in
particular, in the case of thin films with the magnetic field normal to the
film plane (``perpendicular geometry'') \cite{gurevich94,brandt93}. Theory 
\cite{brandt96,gurevich94,brandt93,jooss96a} and experiment \cite
{vlasko,jooss96} show that the magnetic behavior in the perpendicular
geometry has many distinctive features, essentially different from the
parallel geometry, e.g. a more complicated structure of the critical state
and the presence of geometrical barriers \cite{zeldov}.

A number of elegant analytical solutions for the perpendicular geometry (for
strips and disks) describe the Meissner \cite{norris}, the mixed state \cite
{brandt96,brandt93}, and magnetic flux creep \cite{gurevich94}. These
solutions are based on the important ansatz, that one can treat the film as
an infinitesimal thin plane. Then, current distribution related to vortex
bending does not influence the results of the analysis which deals only with
the current density and the vortex displacements averaged over the film
thickness. This approach was very successful in explaining the peculiarities
of the current density and the magnetic induction distribution across the
film plane. However, this approach cannot account for any {\em thickness
dependence} of both persistent current density $j$ \cite
{jooss96,mcelfresh,oneoverd,prozorov97,sheriff97} and magnetic relaxation
rate \cite{prozorov97,sheriff97} in thin films.

Explanation of the observed decrease of $j$ with the increase of the film
thickness $d$ is usually based on the idea that pinning on surfaces {\em %
perpendicular} to the direction of vortices is strong enough and must be
taken into account \cite{jooss96a,jooss96,mcelfresh}. However, as we
demonstrate below, this is not sufficient for understanding the thickness
dependence of the magnetic relaxation rate, which was found to decrease with
the increase of the film thickness \cite{prozorov97,sheriff97}.

Another explanation of the observed thickness dependence of the current
density may be based on collective pinning in a 2D regime, i. e., for
longitudinal correlation length $L$\ larger than the film thickness. This
case is carefully considered in \cite{brandt86}. In a 2D collective pinning
regime, the pinning is stronger for thinner samples. As a result, in this
model both the critical-current density and the creep barrier are larger in
thinner samples, contrary to the experimental results. Also, this scenario
is probably not relevant for the explanation of the experimental data
discussed below, because the thickness of our films $d\geq 800$\AA\ is
larger than $L\approx 40-100$\AA .

In order to understand the experimental results we calculate the current
density and magnetic induction distribution by using the 'two-mode
electrodynamics' theory suggested earlier to explain the AC response in bulk
materials \cite{ST}. The essence of this theory is that two length scales
govern the penetration of fields and currents into type-II superconductors.
The longer scale is of electrodynamic origin and, therefore, is more
universal: it exists, for example, in a superconductor in the Meissner state
(the London penetration depth) or, in a normal conductor (the skin depth).
The shorter scale is related to the vortex-line tension, so it is unique for
a type-II superconductor in the mixed state. This scale was introduced into
the continuous theory of type-II superconductors by Matheiu and Simon \cite
{MS} (see also \cite{brandt,MSS}). When applying the two-mode
electrodynamics to the critical state one may ignore the time variation,
i.e., the two-mode electrodynamics becomes the {\em two-mode electrostatics }%
theory.

Our analysis of a type-II thin superconducting film within the two-mode
electrostatics theory leads to the conclusion that for strong enough bulk
pinning, inhomogeneity of the current density becomes important, even in the
absence of surface pinning, if the film thickness exceeds the Campbell
penetration depth $\lambda _{C}$. Thus, inhomogeneity of the current
distribution throughout the film thickness is a {\em distinctive} and
inevitable feature of the perpendicular film geometry like, for example, the
geometrical barrier \cite{zeldov}. Inhomogeneity of the current distribution
is significantly enhanced if the critical state is supported by the surface
pinning. In this case, most of the current is confined to a layer of a depth
of the order of the intervortex distance, which is usually much smaller than
the London penetration depth $\lambda $ and film thickness. As a result of
this inhomogeneity, the {\it measured} average critical current density
becomes thickness dependent. This current inhomogeneity also causes a
thickness dependence of the magnetic relaxation rate. In the following we
present detailed calculations of the distribution of the current density $j$
and induction field $B$ in thin type-II superconducting film, resulting from
surface and/or bulk pinning. We then introduce the first critical state
model which takes into account the variation in $j$ throughout the film
thickness. Calculations based on this critical state model lead to a
thickness dependence in $j$ and magnetic relaxation rate. These predictions
are compared with the experimental data.

\section{Theory}

\subsection{Equations of electrodynamics for the mixed state in
perpendicular geometry}

Let us consider a thin superconducting strip, infinitely long in the $y$
-direction, with width $2w$ $\left( - w<x<w\right) $ and thickness $2d$ $%
\left( -d<z<d\right) $. External magnetic field $H$ is applied along the $z$
-axis, perpendicular to the film plane. The vortex density $n$ is determined
by the $z$-component $B_z$ of the average magnetic field (magnetic
induction) $\vec B$ in the film: $n=B_z/\Phi _0$. Supercurrent of density $%
I_y\left( x,z\right) $ flows along the $y$-axis resulting in a Lorenz force
in the $x$-direction, and a vortex displacement $u$ along the $x$-axis.

We begin with the electrodynamic equations describing the mixed state of
type-II superconductors in such a geometry. They include the London equation
for the $x$-component of the magnetic field: 
\begin{equation}
B_{x}-\lambda ^{2}\frac{\partial ^{2}B_{x}}{\partial z^{2}}=B_{z}\frac{
\partial u}{\partial z}~,  \label{Lon}
\end{equation}
the Maxwell equation: 
\begin{equation}
{\frac{4\pi }{c}}j_{y}=\frac{\partial B_{x}}{\partial z}-\frac{\partial
B_{z} }{\partial x}~,  \label{Max}
\end{equation}
and the equation of vortex motion: 
\begin{equation}
\eta \frac{\partial u}{\partial t}+ku={\frac{\Phi _{0}}{c}}j_{y}+{\frac{\Phi
_{0}}{4\pi }}H^{*}\frac{\partial ^{2}u}{\partial z^{2}}~,  \label{Vor}
\end{equation}
where 
\begin{equation}
H^{*}=\frac{\Phi _{0}}{4\pi \lambda ^{2}} \ln \frac{a_{0}}{r_{c}}
\label{H-c}
\end{equation}
is a field of order of the first critical field $H_{c1}$, $a_{0}\simeq \sqrt{
\Phi _{0}/B_{z}}$ is the inter-vortex distance, and $r_{c}\sim \xi $ is an
effective vortex core radius. The equation of the vortex motion arises from
the balance among four terms: (i) the friction force proportional to the
friction coefficient $\eta $; (ii) the homogeneous, linear elastic pinning
force $\propto k$ (i. e. assuming small displacements $u$); (iii) the
Lorentz force proportional to the current density $j$; and (iv) the
vortex-line tension force (the last term on the right-hand side of Eq. (\ref
{Vor})), taken from Ref. \cite{ST}.

In the parallel geometry, ($d\rightarrow \infty $), vortices move without
bending so that the $x$-component $B_{x}$ is absent, and the Maxwell
equation becomes: $4\pi j_{y}/c=-\partial B_{z}/\partial x$. Since $B_{z}$
is proportional to the vortex density, this current may be called a {\em %
diffusion current}. The case of the perpendicular geometry, ($d\ll w$), is
essentially different: the diffusion current is small compared to the {\em %
bending current} $\partial B_{x}/\partial z$ (see the estimation below) and
may be neglected for calculation of the distribution throughout the film
thickness (along the $z$-axis). As a result, Eq. (\ref{Vor}) becomes 
\begin{equation}
\eta \frac{\partial u}{\partial t}+ku=\frac{\Phi _{0}}{4\pi }\frac{\partial
B_{x}}{\partial z}+{\frac{\Phi _{0}}{4\pi }}H^{\ast }\frac{\partial ^{2}u}{%
\partial z^{2}}~.  \label{Vor-tr}
\end{equation}
Equations (\ref{Lon}) and (\ref{Vor-tr}) determine the distribution of the
displacement $u(z)$ and of the in-plane magnetic induction $B_{x}(z)$. This
also yields a distribution of the current density $(4\pi
/c)j_{y}(z)=\partial B_{x}(z)/\partial z$. But these equations are still not
closed, since the two components of the magnetic induction, $B_{x}$ and $%
B_{z}$, and current density $j_{y}(z)$ are connected by the Biot-Savart law.
However, neglecting the diffusion current in the Maxwell equation we
separate the problem into two parts: (1) determination of the distribution
of fields and currents along the $z$- axis, taking the total current $%
I_{y}=cB_{x}^{s}/2\pi $ (here $B_{x}^{s}\equiv B_{x}\left( z=d\right) $) and
the perpendicular magnetic-induction component $B_{z}$ as free parameters;
(2) determination of the parameters $I_{y}$ and $B_{z}$ using the Biot-
Savart law. The latter part of the problem (solution of the integral
equation given by the Biot-Savart law) has already been studied carefully in
previous works \cite{brandt96,brandt93}. In the present work we concentrate
on the analysis of the distribution of fields and currents throughout the
film thickness ($z$-dependence).

The accuracy of our approach is determined by the ratio of the diffusion
current $\partial B_{z}/\partial x$ to the bending current $\partial
B_{x}/\partial z$, since we neglect the diffusion current contribution to
the total current. Suppose, as a rough estimation, that $B_{z}\sim B_{x}$%
\cite{vlasko}. Then, the diffusion current density is roughly $\sim I_{y}/w$%
, whereas the bending current density is $\sim I_{y}/d$ \cite
{brandt96,brandt93,vlasko,zeldov}. Thus, the ratio between the diffusion and
the bending current is approximately $d/w\sim 10^{-3}\div 10^{-4}$ for
typical thin films. Note that this condition does not depend on the
magnitude of the critical current and is well satisfied also in typical
single crystals, where $d/w\sim 0.01\div 0.1$. Therefore, the results we
obtain below hold for a wide range of typical samples used in the experiment.

\subsection{Two-mode electrostatics: Two length scales}

Let us consider the static case when vortices do not move, hence there is no
friction. Then, Eq. (\ref{Vor-tr}) becomes 
\begin{equation}
ku={\frac{\Phi _{0}}{4\pi }}\frac{\partial B_{x}}{\partial z}+{\frac{\Phi
_{0}}{4\pi }}H^{\ast }\frac{\partial ^{2}u}{\partial z^{2}}~.  \label{Vor-st}
\end{equation}
Excluding the $B_{x}$ component of the magnetic induction from Eqs. (\ref
{Lon}) and (\ref{Vor-st}) we obtain equation for the vortex displacement: 
\begin{equation}
-{\frac{4\pi k}{\Phi _{0}}}\left( u-\lambda ^{2}\frac{\partial ^{2}u}{%
\partial z^{2}}\right) +(H^{\ast }+B_{z})\frac{\partial ^{2}u}{\partial z^{2}%
}-\lambda ^{2}H^{\ast }\frac{\partial ^{4}u}{\partial z^{4}}=0~.
\label{vort-gen}
\end{equation}

The two length scales which govern distributions over the $z$-axis become
evident if one tries to find a general solution of equation \ref{vort-gen}
in the form $B_{x}\sim u\sim \exp (ipz)$. Then, the dispersion equation for $%
p$ is bi-quadratic and yields two negative values for $p^{2}$. In the limit $%
k\ll 4\pi \lambda ^{2}/\Phi _{0}(H^{\ast }+B_{z})$ (weak bulk pinning): 
\begin{equation}
p_{1}^{2}=-{\frac{1}{\widetilde{\lambda }^{2}}}=-{\frac{1}{\lambda ^{2}}}%
\frac{H^{\ast }+B_{z}}{H^{\ast }}~,  \label{k-1}
\end{equation}
\begin{equation}
p_{2}^{2}=-{\frac{1}{\lambda _{C}^{2}}}=-\frac{4\pi k}{\Phi _{0}(H^{\ast
}+B_{z})}~,  \label{k-2}
\end{equation}

Thus, the distribution along the $z-$axis is characterized by the two length
scales: the Campbell length $\lambda _C$, which is the electrodynamic
length, and length $\widetilde{\lambda }$, given by Eq. (\ref{k-1}), which
is related to $\lambda $ and the vortex-line tension.

\subsection{Current density and field distribution}

In order to determine distribution of currents and fields throughout the
film thickness, one must add the proper boundary conditions to the general
solution of Eq. (\ref{vort-gen}). We look for a solution which is a
superposition of two modes. In particular, for the vortex displacement we
can write: 
\begin{equation}
u(z)=u_0\cosh {\frac z{\lambda _C}}+u_1\cosh {\frac z{\widetilde{\lambda }}}%
~.  \label{u-pin}
\end{equation}
Using Eq. (\ref{Vor-st}) one has for the current density: 
\begin{equation}
{\frac{4\pi }c}j_y=\frac{\partial B_x}{\partial z}\approx B_z\frac{u_0}{%
\lambda _C^2}\cosh {\ \frac z{\lambda _C}}-H^{*}\frac{u_1}{\widetilde{%
\lambda }^2}\cosh {\frac z{\widetilde{\lambda }}}~.  \label{curr}
\end{equation}
The total current is 
\begin{equation}
{\frac{4\pi }c}I_y=2B_x(d)=2B_z{\frac{u_0}{\lambda _C}}\sinh {\frac d{%
\lambda _C}}-2H^{*}{\frac{u_1}{\widetilde{\lambda }}}\sinh {\frac d{%
\widetilde{\lambda }}}~.  \label{curr-tot}
\end{equation}
Equation (\ref{curr-tot}) is in fact a boundary condition imposed on the
amplitudes of two modes, $u_0$ and $u_1$. The second boundary condition is
determined by the strength of the surface pinning. If displacements are
small, the general form of this boundary condition is 
\begin{equation}
\alpha u(\pm d)\pm \left. \frac{\partial u}{\partial z}\right| _{\pm d}=0~,
\label{BC}
\end{equation}
where $\alpha =0$ in the absence of surface pinning and $\alpha \rightarrow
\infty $ in the limit of strong surface pinning. In the following parts of
the section we consider these two limits.

\subsubsection{Surface pinning}

\label{SP}

Let us consider the case of surface pinning in the absence of bulk pinning ($%
k=0$), when the Campbell length $\lambda _{C}\rightarrow \infty $ (see Eq. (%
\ref{k-2})). By ``surface pinning'' we understand pinning due to surface
roughness on the surfaces {\it perpendicular} to the vortex direction. The
surface roughness is assumed to be much smaller than the film thickness $d$.
By substituting $\lambda _{C}\rightarrow \infty $ in the general solution
Eq. (\ref{u-pin}), we derive the displacement for surface pinning: 
\begin{equation}
u(z)=u_{0}+u_{1}\cosh {\frac{z}{\widetilde{\lambda }}}~,  \label{u-z}
\end{equation}
where $u_{0}$ and $u_{1}$ are constants, which can be determined from the
boundary conditions Eqs. (\ref{curr-tot}) and (\ref{BC}). Note, however,
that $u_{0}$ is not important in the case of surface pinning, because the
constant $u_{0}$ does not affect distributions of currents and fields.

The magnetic field $B_{x}$ is obtained from Eq. (\ref{Vor-st}): 
\begin{equation}
B_{x}(z)=-H^{\ast }{\frac{u_{1}}{\widetilde{\lambda }}}\sinh {\frac{z}{%
\widetilde{\lambda }}},  \label{B-x}
\end{equation}
and the current is determined from the Maxwell equation (\ref{Max})
neglecting the diffusion current: 
\begin{equation}
j_{y}=-{\frac{c}{4\pi }}H^{\ast }{\frac{u_{1}}{\widetilde{\lambda }^{2}}}%
\cosh {\frac{z}{\widetilde{\lambda }}}.  \label{j-y}
\end{equation}

It is important to note that the characteristic length $\widetilde{\lambda }$
, which varies between the London penetration length $\lambda $ and the
inter-vortex distance $a_0\sim \sqrt{\Phi _0/B_z}$, is much smaller than $%
\lambda $ for a dense vortex array, $B_z\gg H^{*}$. Taking into account that
usually thin films have thickness less or equal to $2\lambda $, the effect
of the vortex bending due to surface pinning may be very important: most of
the current is confined to a thin surface layer of width $\widetilde{\lambda 
}$.

The current density on the surface is $j_{s}\equiv j_{y}\left( z=d\right) =$ 
$-{\ \ }\left( c/4\pi \right) H^{\ast }\left( u_{1}/\tilde{\lambda}%
^{2}\right) \cosh \left( d/\tilde{\lambda}\right) $. Thus, 
\begin{equation}
u_{1}=-\frac{4\pi }{c}\frac{\widetilde{\lambda }^{2}j_{s}}{H^{\ast }\cosh {%
\frac{d}{\tilde{\lambda}}}}.  \label{u1}
\end{equation}

\noindent The total current integrated over the film thickness $2d$ is: 
\begin{equation}
I_{y}=\int_{-d}^{d}j_{y}(z)dz=-{\frac{c}{2\pi }}H^{\ast }{\frac{u_{1}}{%
\tilde{\lambda}}}\sinh {\frac{d}{\tilde{\lambda}}=2}\widetilde{{\lambda }}%
j_{s}\tanh \frac{d}{\widetilde{\lambda }}.  \label{J-y}
\end{equation}
Thus, the {\em average} current density $j_{a}\equiv I_{y}/2d$ - the
quantity derived in the experiment - decreases with thickness as 
\begin{equation}
j_{a}=j_{s}\frac{\widetilde{\lambda }}{d}\tanh \frac{d}{\widetilde{\lambda }}%
,  \label{ja-surf}
\end{equation}
yielding $j_{a}=j_{s}\widetilde{\lambda }/d$ for $\widetilde{\lambda }/d<<1$
as found experimentally \cite{prozorov97}.

The field and the current distribution over the film thickness are: 
\begin{equation}
j_{y}(z)=\frac{I_{y}}{2\widetilde{\lambda }}\frac{\cosh {\frac{z}{\tilde{%
\lambda}}}}{\sinh {\frac{d}{\tilde{\lambda}}}}=j_{s}\frac{\cosh {\frac{z}{%
\tilde{\lambda}}}}{\cosh {\frac{d}{\tilde{\lambda}}}},  \label{j-y-J}
\end{equation}
\begin{equation}
B_{x}(z)=\frac{2\pi }{c}I_{y}\frac{\sinh {\frac{z}{\widetilde{\lambda }}}}{%
\sinh {\ \frac{d}{\widetilde{\lambda }}}}=\frac{4\pi }{c}j_{s}\widetilde{%
\lambda }\frac{\sinh {\frac{z}{\widetilde{\lambda }}}}{\cosh {\frac{d}{%
\widetilde{\lambda }}}}.  \label{B-x-J}
\end{equation}
Thus, the current penetrates into a small depth $\tilde{\lambda}$ and is
exponentially small in the bulk beyond this length.

\subsubsection{Bulk pinning}

A remarkable feature of the perpendicular geometry is that, even in the
absence of surface pinning, vortices are bent. This is in striking contrast
with the parallel geometry where the diffusion current distribution is
homogeneous along the direction of vortices and, therefore, does not bend
them. Absence of surface pinning means that at the surface $\partial
u/\partial z=0$ (a vortex is perpendicular to an ideal surface). This yields
the relation between $u_0$ and $u_1$ [see Eq. (\ref{u-pin})]:

\[
u_1=-u_0\frac{\widetilde{\lambda }}{\lambda _C}\frac{\sinh \frac z{\lambda
_C }}{\sinh \frac z{\widetilde{\lambda }}} 
\]

Then, Eq. (\ref{curr-tot}) becomes

\begin{equation}
{\frac{4\pi }c}I_y=2(B_z+H^{*}){\frac{u_0}{\lambda _C}}\sinh {\frac d{
\lambda _C}}~.  \label{curr-tot1}
\end{equation}

\noindent The current distribution is 
\begin{equation}
j_y(z)=I_y\left( \frac 1{2\lambda _C}\frac{B_z}{H^{*}+B_z}\frac{\cosh {\frac %
z{\lambda _C}}}{\sinh {\frac d{\lambda _C}}}+{\frac 1{2\widetilde{\lambda }}}
\frac{H^{*}}{H^{*}+B_z}\frac{\cosh {\frac z{\widetilde{\lambda }}}}{\sinh {\ 
\frac d{\widetilde{\lambda }}}}\right) ~.  \label{cur-dist}
\end{equation}
In the limit $d\ll \lambda _C$ Eq. (\ref{cur-dist}) yields 
\begin{equation}
j_y(z)=I_y\left( \frac 1{2d}\frac{B_z}{H^{*}+B_z}+{\frac 1{2\tilde \lambda }}
\frac{H^{*}}{H^{*}+B_z}\frac{\cosh {\frac z{\widetilde{\lambda }}}}{\sinh {\ 
\frac d{\tilde \lambda }}}\right) ~.  \label{d-sm}
\end{equation}

Another interesting case is that of the dense vortex array, $B_{z}\gg
H^{\ast }$: 
\begin{equation}
j_{y}(z)=\frac{I_{y}}{2\lambda _{C}}\frac{\cosh {\frac{z}{\lambda _{C}}}}{%
\sinh {\ \frac{d}{\lambda _{C}}}}=j_{s}\frac{\cosh {\frac{z}{\lambda _{C}}}}{%
\cosh {\frac{d}{\lambda _{C}}}}~,  \label{cur-Camp}
\end{equation}
where again $j_{s}$ is the current density on the film surface. Remarkably,
current density is inhomogeneous even in the absence of surface pinning. We
illustrate this in Fig.1, where we plot $j_{y}\left( z\right) /j_{b}$ vs. $%
z/d$ at different ratios $d/\lambda _{C}$. ``Uniform'' bulk current density $%
j_{b}=I_{y}/2d$ corresponds to the limit $d/\lambda _{C}=0$. Physically,
such current profiles reflect Meissner screening of the in-plane component $%
B_{x}$ of the self-field.

For the average current density we have 
\begin{equation}
j_a=j_s\frac{\lambda _C}d\tanh \frac d{\lambda _C}~,  \label{ja-bulk}
\end{equation}
which is similar to the case of the surface pinning, Eq. (\ref{ja-surf}),
with $\widetilde{\lambda }$ replaced by $\lambda _C$.

Thus, in the perpendicular geometry, the current distribution is strongly
inhomogeneous: the whole current is confined to a narrow surface layer of
width $\widetilde{\lambda }$ (surface pinning), or $\lambda _C$ (bulk
pinning).

\subsection{Critical state}

In the theory given in the previous sections we have assumed that currents
and vortex displacements are small. In this section we deal with the
critical state when the current density equals its critical value $j_c$. Let
us consider how it can affect our picture, derived in the previous sections
for small currents.

\subsubsection{Surface pinning}

If vortices are pinned only at the surface, the value of the critical
current depends on the profile of the surface, and one may not use the
linear boundary condition imposed on the vortex displacement, Eq. (\ref{BC}
). However, the $z$-independent vortex displacement $u_0$ does not influence
the current density and field distribution in the bulk as shown in Sec. \ref
{SP} (see Eqs. (\ref{B-x}) and (\ref{j-y})). Therefore the bulk current
density and field distribution derived from our linear analysis can be used
even for the critical state.

\subsubsection{Bulk pinning}

\label{BP}

In this case our theory must be modified for the critical state. In
particular, for large currents the bulk pinning force becomes nonlinear and,
as a result, the current and field penetration is not described by simple
exponential modes. Formally, this nonlinearity may be incorporated into our
theory assuming a $u$ - dependent pinning constant $k$, thus allowing $k$ to
vary along the vortex line. As an example, let us consider the case of
strongly localized pinning force when the vortex is pinned by a potential
well of a small radius $r_{d}$ like that sketched in Fig.2: the vortex
energy per unit length (vortex-line tension) is given by $\varepsilon $ for
vortex line segments outside the potential well and by $\varepsilon _{0}$
for segments inside the well. Thus, the pinning energy per unit length is $%
\varepsilon -\varepsilon _{0}$. In fact, such a potential well model may
describe pinning of vortices by, for example, one-dimensional columnar
defects or planar defects, such as twin or grain boundaries\cite
{nucl,Sonin95}. The latter is relevant in thin films obtained by usual
method of laser ablation.{\large \ }Therefore, we can also use such a
pinning potential as a rough qualitative model for typical types of pinning
sites, in order to illustrate the effect of bulk pinning on the current
density distribution and the rate of magnetic relaxation in thin films.

If the current distribution were uniform, such a potential well would keep
the vortex pinned until the current density $j_{y}$ exceeds the critical
value $c(\varepsilon -\varepsilon _{0})/\Phi _{0}r_{d}$. The escape of the
trapped vortex line from the potential well occurs via formation of the
un-trapped circular segment of the vortex line (see Fig.3(a)). In this case,
both the critical-current density and the energy barrier for vortex
depinning do not depend on film thickness \cite{nucl}.

But, in perpendicular geometry the current distribution is not homogeneous.
In order to find it for the critical state, we may use the following
approach. The vortex line consists of the trapped and untrapped segments as
shown in Fig.3(b). The untrapped segment is beyond the potential well,
therefore there is no bulk pinning force acting on it. This means that the
shape of this segment is described by Eq. (\ref{Vor-st}) with $k=0$.
Applying the theory of Sec. \ref{SP}, one obtains that the total current $%
I_{y}=\int_{-d}^{d}j_{y}(z)dz$ is concentrated near the film surfaces within
a narrow surface layer of width $\widetilde{\lambda }$. Inside the surface
layer the vortex line is curved, but has a straight segment of length $L$
outside the layer, as illustrated in Fig.3(b). As for the vortex-line
segment trapped by the potential well, we assume that it is straight and
vertical, neglecting its possible displacements inside the potential well.
Formally speaking, our approach introduces a non-homogeneous bulk-pinning
constant $k$ assuming that $k=0$ for the untrapped segment and $k=\infty $
for the trapped one. The energy of the vortex line in this state is
determined by the line tensions ($\varepsilon $ and $\varepsilon _{0}$) and
is given by 
\begin{equation}
E=2\varepsilon {\frac{L}{\cos \alpha }}-2\varepsilon _{0}L-2{\frac{\Phi _{0}%
}{c}}I_{y}L\tan \alpha =2L\tan \alpha \left( \varepsilon \sin \alpha -{\frac{%
\Phi _{0}}{c}}I_{y}\right) ~,  \label{E-J}
\end{equation}
where the contact angle $\alpha $ is determined by the balance of the
line-tension forces at the point where the vortex line meets the line
defect: 
\begin{equation}
\cos \alpha =\frac{\varepsilon _{0}}{\varepsilon }~.  \label{alpha}
\end{equation}

\section{Magnetic relaxation}

\label{relaxation}

We now discuss the effect of current density distribution on the thickness
dependence of magnetic relaxation. We first show below, that uniform current
density cannot explain the experimentally observed thickness dependence. We
also show that inhomogeneous current density distribution, resulting from
the surface pinning only, cannot explain the experimental data too. We
demonstrate that only presence of a bulk pinning and the resulting current
inhomogeneity may lead to an accelerated relaxation in thinner films. We
finally discuss the general case when both bulk and surface pinning are
present.

As pointed out above, if the current distribution is uniform throughout the
film thickness, a trapped vortex may escape from the potential well (Fig.2)
via formation of a circular segment of the vortex line (Fig.3(a)), with the
energy 
\begin{equation}
E=\varepsilon L-\varepsilon _{0}L_{0}-{\frac{\Phi _{0}}{c}}j_{y}S,
\label{E-par}
\end{equation}
where $L$ and $L_{0}$ are the lengths of the vortex line segment before and
after formation of the loop, $S$ is the area of the loop \cite{Sonin95,nucl}%
. If the loop is a circular arc of the radius $R$ and the angle $2\alpha $
(Fig.3(a)), then $L_{0}=2R\sin \alpha $, $L=2R\alpha $, and $S={\frac{1}{2}}%
R^{2}(2\alpha -\sin 2\alpha )$, where the contact angle $\alpha $ is given
by Eq. (\ref{alpha}). Then, 
\begin{equation}
E=2R(\varepsilon \alpha -\varepsilon _{0}\sin \alpha )-\frac{\Phi _{0}}{2c}%
j_{y}R^{2}(2\alpha -\sin 2\alpha )=(2\alpha -\sin 2\alpha )\left(
\varepsilon R-{\ \frac{\Phi _{0}}{2c}}j_{y}R^{2}\right) .  \label{E-par-fin}
\end{equation}
The height of the barrier is determined by the maximum energy at $%
R_{c}=\varepsilon c/\Phi _{0}j_{y}$: 
\begin{equation}
E_{b}=(2\alpha -\sin 2\alpha )\frac{\varepsilon ^{2}c}{2\Phi _{0}j_{y}}.
\label{bar}
\end{equation}
As one might expect, this barrier and consequently the relaxation rate do
not depend on the film thickness. We stress that this estimation is valid
only for $d>R_{c}$. If $d<R_{c}$\ the energy barrier is obtained from Eq. (%
\ref{E-par-fin}) by substituting $R=d.$ This case of uniform current,
however, leads to a thickness independent current density, and therefore
cannot describe the experimental data.

\subsection{Surface pinning}

In this case, the whole current is confined to the surface layer of width $%
\widetilde{\lambda }$. It is apparent from Eq.\ (\ref{k-2}) that for typical
experimental fields ($\sim 1T$) $\widetilde{\lambda }$ is smaller than the
film thickness. This means that current flows mostly in a thin surface
layer. Thus, all creep parameters, including the creep barrier, are governed
by the total current $I_y$, and not by the average current density $I_y/2d$.
Then, apparently, the critical current density and the creep barrier are
larger for thinner films, similar to the case of the collective- pinning
effect mentioned above. Thus, also this scenario cannot explain the observed
accelerated relaxations in the thinner films.

\subsection{Short-range bulk pinning}

\label{creep-bulk}

Let us consider the relaxation process for a critical state supported by the
short-range pinning force discussed in Sec. \ref{BP}. The energy $E$ of the
vortex line is given by Eq. (\ref{E-J}). The average critical current
density corresponds to $E=0$ and is inversely proportional to the film
thickness [see also Eq. (\ref{ja-bulk})]: 
\begin{equation}
j_{c}=\frac{I_{c}}{2d}=\frac{c\varepsilon }{2d\Phi _{0}}\sin \alpha ~.
\label{j-c}
\end{equation}
The energy barrier is given by the maximum energy at $d=L+\tilde{\lambda}%
\approx L$ when the whole vortex line has left the potential well
(Fig.4(a)): 
\begin{equation}
E_{b}=\tan \alpha \left( 2d\varepsilon \sin \alpha -4d^{2}{\frac{\Phi _{0}}{c%
}}j_{a}\right) ~,  \label{bar-J}
\end{equation}
where $j_{a}=I_{y}/2d$ is the average current density. If $%
j_{c}>j_{a}>j_{c}/2$, then $\partial E_{b}/\partial d<0$, i.e., the barrier
is larger for thinner films. But, for $j_{a}<j_{c}/2$ the derivative $%
\partial E_{b}/\partial d>0$ , and the barrier {\em increases} with the
increase of the film thickness. Thus, under this condition ($j_{a}<j_{c}/2$)
the magnetic relaxation rate is larger in the thinner samples.

The above analysis did not take into account the possibility for dense
defects. By ``dense'' we mean that the distance $r_{i}$ from the neighbor
potential well is less than $d\tan \alpha $ (see Fig.4(b)). In this case the
maximal energy (the barrier peak) is smaller than the barrier calculated in
Eq. (\ref{bar-J}). Then the barrier energy is given by 
\begin{equation}
E_{b}=r_{i}\left( 2\varepsilon \sin \alpha -4d{\frac{\Phi _{0}}{c}}%
j_{a}\right)  \label{bar-def}
\end{equation}
In this case $\partial E_{b}/\partial d<0$ and the energy barrier for
thinner films is always larger. Therefore one can see faster relaxation in
thinner films only if the films are so thin that $d<r_{i}/\tan \alpha $ and
the energy barrier is given by Eq. (\ref{bar-J}). From the experimental
results shown below we infer that the average distance between effective
defects $r_{i}\geq 1000\ \AA $\ in agreement with direct measurements using
atomic force microscopy.

To conclude, if the average current density in thin films becomes small
enough compared to the original critical current density and if the films
are thin enough, the relaxation {\em at the same average persistent current}
is predicted to be faster for the thinner films.

\subsection{General case}

\label{creep-general}

In the simplified picture of the critical-state relaxation outlined in the
previous subsection, the total current was concentrated within a very thin
layer of the width $\widetilde{\lambda }$. It was based on the assumption
that the pinning force disappears when the vortex line leaves the small-size
potential well, whereas inside the potential well the pinning force is very
strong. As a result, outside the thin surface layers of the width $%
\widetilde{\lambda }$ the vortex line consists of two straight segments
(Figs.3(b) and 4). In the general case, the distribution of the pinning
force may be smoother and the shape of a vortex line is more complicated. In
addition, interactions between the vortices may modify the barrier for flux
creep as well. However, the tendency must be the same: the current confined
in a narrow surface layer drives the end of a vortex line away from the
potential well to the regions where the pinning force is weaker and the
vortex line is quite straight with the length proportional to thickness of
the film if the latter is thin enough. Therefore, the barrier height for the
vortex jump is smaller for smaller $d$.

We also note that we do not consider an anisotropic case and limit our
discussion to isotropic samples. The effect of anisotropy on the barrier
height was considered in details in Ref.\cite{nucl} In the presence of
anisotropy the circular loop becomes elliptic and the vortex-line tension $%
\varepsilon $\ must be replaced by some combination of vortex-line tensions
for different crystal directions. These quantitative modifications are not
essential for our qualitative analysis.

Our scenario assumes that the current is concentrated near the film
surfaces. In general, width of the current layer may vary from $\widetilde{%
\lambda }$ to effective Campbell length $\lambda _{C}$. One may then expect
a {\em non-monotonous} thickness dependence when $\lambda _{C}$ is
comparable with $d$. As we see, the Campbell length is an important quantity
in determining whether current density inhomogeneity must be taken into
account or not (in the absence of the surface pinning). The length $\lambda
_{C}$ can be estimated from the micro-wave experiments: according to
Golosovskii {\it et al.}\cite{golosovskii} $\lambda _{C}\simeq 1000\sqrt{H}\
\AA $, where the field $H$ is measured in $Tesla$. For $H\simeq 0.2$ $T$
this results in $\lambda _{C}\approx 450\ \AA $\ or $2\lambda _{C}\approx
900\ \AA $, which has to be compared with the film thickness.

\section{Comparison with the experiment}

A decrease of the measured current density with an increase of the film
thickness is reported in numerous experimental works \cite
{jooss96,oneoverd,prozorov97,sheriff97}. This is consistent with the
predictions given above for either surface or/and bulk pinning. Both pinning
mechanisms predict similar $1/d$ dependence of $j$ and it is, therefore,
impossible to distinguish between surface and bulk pinning in this type of
measurements. Only the additional information from the thickness dependence
of the relaxation rate allows the drawing of some conclusions about the
pinning mechanisms.

Magnetic relaxation measurements in films of different thickness are
discussed in detail in \cite{prozorov97,sheriff97}. Using excerpts from the
data reported there we demonstrate an agreement of these data with our
theory.

Measurements were conducted on four $5\times 5$ $mm^{2}$ $%
YBa_{2}Cu_{3}O_{7-\delta }$ films of thickness $2d=800,\ 1000,\ 2000$ and $%
3000\ \AA $, prepared by the laser ablation technique on $SrTiO_{3}$
substrates \cite{koren}. All samples had $T_{c}\approx 89\ K$. The
morphology of the samples was examined by atomic-force microscopy (AFM)
technique and was found to be similar: the average grain size $(1-50)\times
10^{2}\ \AA $\ and intergrain distance $50\ \AA $ (For typical AFM picture
of our samples, see Fig. 1(c) in \cite{sheriff97}).{\large \ }The magnetic
moment was measured as a function of field, temperature and time, using a 
{\it Quantum Design} SQUID magnetometer.

The {\em average} persistent current density was extracted from the magnetic
hysteresis loops using the Bean model adapted for our case: $j_{a}\left[
A/cm^{2}\right] =30M/da^{3}$, where $M\left[ emu\right] $ is the
irreversible magnetic moment, $d\left[ cm\right] $ is a half of the film
thickness and $a=0.5\ cm$ is the lateral dimension. Fig.5 shows the
persistent current density $j$ at $T=5\ K$ as a function of the applied
magnetic field $H$. Apparently, $j$ is larger in thinner films. The same
trend is found at all temperatures. These observations are in good agreement
with Eqs. (\ref{ja-surf}) and (\ref{ja-bulk}). We note, however, that since
the value of $j_{s}$ is not known, we cannot point out the dominance of pure
surface, pure bulk or a mixed type of pinning. On the other hand it is
unlikely that the observed thickness dependence is due to changes in the
density of pinning centers with thickness, since the films' morphology is
similar for all of our samples. This is further indirectly confirmed by the
relaxation measurements. The decrease of current density due to increase of
a mean grain size in thicker films would simultaneously result in faster
relaxation, contrary to our observations.

Fig.6 shows typical relaxation curves at $H=0.2\ T$ (ramped down from $1\
Tesla$) measured in films of different thickness. The interesting and
unexpected feature is that curves cross, i. e., the relaxation is faster in
thinner films. This is further illustrated in Fig.7 where $j$ vs. $d$ is
plotted at different times. At the beginning of the relaxation process, the
average current density in the thinner films is larger. However, in the
thinner films, the current density decreases much faster than in the thicker
ones; as a result $j_{a}$ exhibits a non-monotonous dependence on thickness
at later times, as shown in Fig.7. The faster relaxation in thinner films is
in qualitative agreement with our results, discussed in Sec. \ref{relaxation}%
, in particular in subsections \ref{creep-bulk} and \ref{creep-general}.
There, we find that such acceleration of the relaxation in thinner films may
be understood only if we consider inhomogeneous bulk current density. In
reality, it is very probable that {\em both} surface and bulk pinning
mechanisms lead to inhomogeneous current density with a characteristic
length scale in between the short (surface pinning) length $\widetilde{%
\lambda }$ and the larger Campbell length.

\section{Summary and Conclusions}

Based on the two mode electrostatics approach we built a consistent theory
of the critical state in thin type-II superconducting films {\em throughout
the film thickness}. We show that, irrespective of the pinning mechanism,
current density is always larger near the surface, and decays over a
characteristic length scale, which is in between $\widetilde{\lambda }$ (of
order of the intervortex distance) and the Campbell length $\lambda _C$. The
length scale $\widetilde{\lambda }$ is determined by the (finite) vortex
tension and by the boundary conditions which force vortices to be
perpendicular to the surface of superconductor, whereas the Campbell length $%
\lambda _C$ is determined by bulk pinning potential.

Following this novel physical picture we conclude that:

\begin{itemize}
\item  Current density and magnetic induction in thin films in perpendicular
field are highly inhomogeneous throughout the film thickness. Surface
pinning significantly enhances these inhomogeneities.

\item  Average current density decreases with the increase of film thickness
approximately as $1/d$.

\item  Magnetic relaxation is {\em slower} in thinner films in the following
cases: (1) In the absence of bulk pinning, i.e., only surface pinning is
effective. (2) In the presence of bulk pinning, provided that the ratio
between thickness and distance between neighboring defects is above a
certain threshold ($d/a\sim 1)$.

\item  Magnetic relaxation is {\em faster} in thinner films only if bulk
pinning is effective and the ratio $d/a$ is below this threshold.
\end{itemize}

In the experimental data presented here the measured average current $j_{a}$
decreases with the increase of film thickness as predicted, and the
relaxation rate is larger for the thinner films, suggesting that $d/a\sim 1$%
, and the effective distance between defects $\geq 1000\ \AA $.

{\sl Acknowledgments:} We thank V. Vinokur, E. H. Brandt, L. Burlachkov, E.
Zeldov and M. Golosovskii for useful discussions. This work was supported in
part by a grant from the Israel Science Foundation and the Heinrich Hertz
Minerva Center for High Temperature Superconductivity, and in part by the
German - Israeli Foundation under Grant $128-3-3/95$. R. P. acknowledges a
support from the Clore Foundations. E. B. S. acknowledges a support by the
Lady Davis Grant and thanks the Racah Institute of the Hebrew University for
hospitality.

\bigskip

\newpage {}{\LARGE Figure captions}

Fig.1 Current density distribution vs. normalized depth $z/d$ for the
indicated $d/\lambda _{c}$ ratios. The current distribution becomes more
inhomogeneous as the ratio $d/\lambda _{c}$ increases.

Fig.2 Vortex energy (per unit length) in the vicinity of the pinning center
of radius $r_{d}$.

Fig.3 (a)Vortex depinning by a uniform current. (b) Simple scenario of
vortex depinning by nonhomogeneous current flowing in a layer $\widetilde{%
\lambda }$.

Fig.4 (a) Barrier maximum configuration in the case of a dilute pinning
centers. (b) Barrier maximum configuration in the case of the dense deffects.

Fig.5 Average persistent current density $j_{a}$ as a function of magnetic
field at $T=5\ K$ for films of different thickness.

Fig.6 Time evolution of the average persistent current density $j_{a}$ at $%
T=75\ K$ for films of different thickness.

Fig.7 Thickness dependence of the average persistent current density $j_{a}$
at $T=75\ K$ taken at different times.

\end{document}